\documentstyle[11pt,epsf]{article}

\newcommand{\beq}{\begin{equation}}
\newcommand{\eeq}{\end{equation}}
\newcommand{\beqa}{\begin{eqnarray}}
\newcommand{\eeqa}{\end{eqnarray}}

\begin{document}

\begin{titlepage}
\def\thepage {}        

\title{A Comment on the Zero Temperature Chiral \\
Phase Transition in $SU(N)$ Gauge Theories}

\author{
R. Sekhar Chivukula\thanks{e-mail addresses: sekhar@bu.edu}\\
Department of Physics, Boston University, \\
590 Commonwealth Ave., Boston MA  02215}

\date{December 5, 1996}

\maketitle

\bigskip
\begin{picture}(0,0)(0,0)
\put(295,250){BUHEP-96-45}
\put(295,235){hep-ph/9612267}
\end{picture}
\vspace{24pt}

\begin{abstract}

  Recently Appelquist, Terning, and Wijewardhana investigated the zero
  temperature chiral phase transition in $SU(N)$ gauge theory as the
  number of fermions $N_f$ is varied. They argued that there is a
  critical number of fermions $N^c_f$, above which there is no chiral
  symmetry breaking and below which chiral symmetry breaking and
  confinement set in.  They further argued that that the transition is
  not second order even though the order parameter for chiral symmetry
  breaking vanishes continuously as $N_f$ approaches $N^c_f$ on the
  broken side.  In this note I propose a simple physical picture for the
  spectrum of states as $N_f$ approaches $ N^c_f$ from below ({\it i.e.}
  on the broken side) and argue that this picture predicts very
  different and {\it non-universal} behavior than is the case in an
  ordinary second order phase transition. In this way the transition can
  be {\it continuous} without behaving conventionally. I further argue
  that this feature results from the (presumed) existence of an infrared
  Banks-Zaks fixed point of the gauge coupling in the neighborhood of
  the chiral transition and therefore depends on the long-distance
  nature of the non-abelian gauge force.

\pagestyle{empty}
\end{abstract}
\end{titlepage}

\setcounter{section}{1}

Recently Appelquist, Terning, and Wijewardhana \cite{atw} have
investigated the zero temperature chiral phase transition in $SU(N)$
gauge theory as the number of massless Dirac fermions $N_f$ is varied.
To second order, the beta-function of such a theory is
given by
\beq
\mu{{\partial}\over{\partial \mu}} \alpha(\mu) = \beta(\alpha)
\equiv -b\, \alpha^2(\mu) -c\, \alpha^3(\mu)-d\, \alpha^4(\mu) - ...~,
\label{beta}
\eeq
where
\beq
b = {{1}\over{6 \pi}} \left( 11 N - 2 N_f\right)
\label{b}
\eeq
\beq
c = {{1}\over {24  \pi^2}} \left(34 N^2 - 10  N N_f - 3{{N^2 -
1}\over{N}} N_f\right)~.
\label{c}
\eeq
For a small number of flavors the theory is asymptotically free and one
expects QCD-like behavior, with confinement and with the chiral
$SU(N_f)_L \times SU(N_f)_R$ symmetry broken to its vectorial subgroup.
If the number of flavors is large enough (in perturbation theory,
greater than $11N/2$) asymptotic freedom (and hence chiral symmetry
breaking and confinement) is lost.  For a range of $N_f$ less than
$11N/2$ the first term in the $\beta$-function is negative and the
theory is asymptotically free, but there appears (in perturbation
theory) to be a nontrivial infrared (Banks-Zaks) fixed point \cite{bz}
because the second term in the $\beta$-function is {\it positive}.  For
$N_f$ just slightly less than $11N/2$, this fixed point $\alpha^*$ is at
weak coupling and the analysis is self-consistent. As $N_f$ is lowered
further, the fixed point moves to {\it larger} coupling.

In vector-like gauge theories an analysis of the gap equation \cite{gap, cjt}
suggests that, in a theory with an approximately constant coupling,
chiral symmetry breaking occurs only if the coupling $\alpha$ exceeds a
critical value
\beq
\alpha_c \equiv {{ \pi }\over{3 \, C_2(R)}}= {{2 \pi \,
N}\over{3\left(N^2-1\right)}}~.
\label{alphacrit}
\eeq
The authors of \cite{atw} suggested that the Banks-Zaks fixed point
\cite{bz} persists to large coupling, and that the chiral-symmetry
breaking transition in $N_f$ should be associated with the point where
$\alpha^* = \alpha_c$. They thereby estimated that
\beq
N_f^c = N \left({{100N^2 -66}\over{25 N^2 -15}}\right).
\label{Ncrit}
\eeq

Furthermore, Appelquist, Terning and Wijewardhana suggested that the
nature of this phase transition is peculiar.  Based on a gap-equation
analysis \cite{miransky} of the the dynamical mass $\Sigma(p)$ of the
fermions in the broken phase, they argued that the order parameter for
chiral symmetry breaking (which is proportional to $\Sigma(0)$)
\beq
\Sigma(0) \approx \Lambda
\exp\left({{- \pi}\over{\sqrt{{{\alpha_*}\over{\alpha_c}} -1}}}\right) ~.
\label{critical}
\eeq 
goes to zero continuously\footnote{ For momenta below $\Sigma(0)$, the
  fermions can be integrated out and no longer contribute to the
  $\beta$-function.  Therefore, strictly speaking, for any $N_f < N^c_f$
  there is no fixed-point. Nonetheless, for $N_f$ close to $N^c_f$ the
  coupling remains close to $\alpha^*$ for a large range of momenta, in
  a manner reminiscent of ``walking technicolor'' \cite{walking}.} as
$N_f \to (N^c_f)^-$ (and $\alpha^* \to \alpha_c^+$).  Here the
high-energy scale $\Lambda$ represents the scale at which the coupling
is far enough from the fixed point value to begin to run \cite{atw}.

On the other hand, based of an analysis of the Bethe-Salpeter equation
for the quark-antiquark scattering amplitude, Appelquist, Terning, and
Wijewardhana argue that in the symmetric phase (close to the transition,
when $N_f$ is just above $N^c_f$) there are no light scalar resonances.
Indeed, since the theory with $N_f \stackrel{>}{\sim} N^c_f$ is presumed
to be in a conformally-invariant ``non-abelian coulomb'' phase
\cite{bz}, it cannot have {\it any} isolated states.  They then
concluded that the phase transition is not second order {\it even though
  the order parameter changes continuously.}

In this note I propose a simple physical picture for the spectrum of
states as $N_f$ approaches $ N^c_f$ from below ({\it i.e.} on the broken
side) and argue that this picture predicts very different and {\it
  non-universal} behavior than is the case in an ordinary second order
phase transition. In this way the transition can be {\it continuous}
without behaving conventionally \cite{yamawaki}.  I further argue that
this feature results from the (presumed) existence of an infrared
Banks-Zaks fixed point of the gauge coupling in the neighborhood of the
chiral transition and therefore depends on the long-distance nature of
the non-abelian gauge force.

\begin{figure}[tbp]
\begin{minipage}{11cm}       
\centering
\epsfysize=2in
\hspace*{0in}
\epsffile{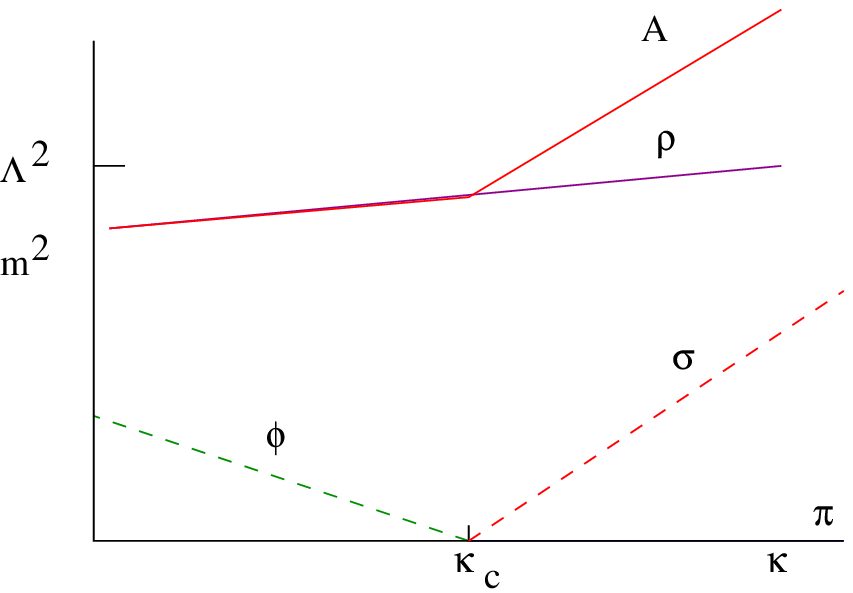}
\caption{Spectrum of bosonic excitations in NJL model for couplings
close to $\kappa_c$. Chiral symmetry is broken for $\kappa >
\kappa_c$, and preserved for $\kappa < \kappa_c$. The $\sigma$ and $\pi$
in the broken phase combine to be the scalar $\phi$ multiplet in the
unbroken phase. The vector and axial vector resonances form
one chiral multiplet, and are therefore degenerate, in the
unbroken phase.}
\label{Fig1}
\end{minipage}
\end{figure}

Before moving to the case of a long-range force, let us review the familiar
behavior of the chiral phase transition in a model with short-range
interactions, the NJL model \cite{njl}. Here the fundamental interactions are
modeled by chirally-invariant local four-fermion operators 
\beq
{\cal L} = -{{4\pi
\kappa}\over{\Lambda^2}}\left[\overline{\psi}\gamma_\mu {{\lambda^a}\over{2}}
\psi \right]^2~,
\label{L4t}
\eeq
where the $\lambda^a/2$ are the generators of $SU(N)$ ``color'' and the
$\psi$ are the $N_f$ flavors of fermions. This interaction is attractive
in the chiral symmetry breaking channel, but is suppressed by a (large)
energy scale $\Lambda$. In the limit where the NJL coupling $\kappa$ is
small, chiral symmetry remains unbroken. When the coupling is large, the
chiral symmetry breaking scale (as characterized by the value of the
$F$-constant --- the analog of $f_\pi$ in QCD --- or by $\Sigma$, the
momentum-independent dynamical mass of the fermion) is of order
$\Lambda$.  There is a critical value $\kappa_c$, estimated to be
$\pi/3$ in the large-$N$ limit, below which chiral symmetry is unbroken
and above which it is broken.  If the transition between these two
regimes is {\it smooth}, as it is in the gap-equation in the
fermion-bubble approximation, the dynamical mass of the fermion goes
smoothly to zero as $\kappa \to (\kappa_c)^+$, and remains identically
zero for $\kappa \le \kappa_c$.

In the case of the Nambu-Jona-Lasinio model, the spectrum of bosonic
excitations \cite{meissner} is shown in Figure 1.  Note that, aside from
the light scalar-multiplet $\phi$ on the unbroken side and the $\sigma$
and Goldstone bosons $\pi$ on the broken side, {\it all} other
``excitations'' have a mass of order $\Lambda$! The reason for this is
that the intrinsic physical scale of the interactions is {\it always} of
order $\Lambda$. Near $\kappa = \kappa_c$, the scalar states are {\it
  anomalously light} due to the ``fine-tuning'' \cite{thooft} of the NJL
interaction.

\begin{figure}[tbp]
\begin{minipage}{11cm}       
\centering
\epsfysize=2in
\hspace*{0in}
\epsffile{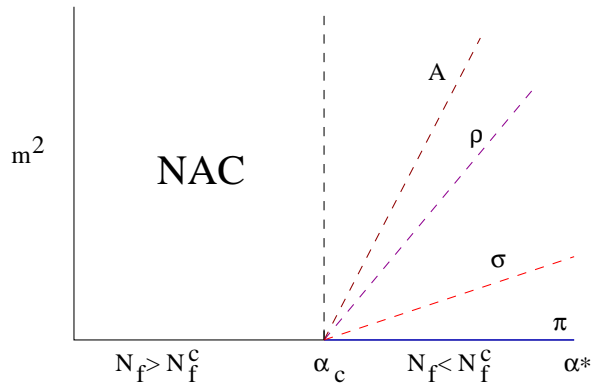}
\caption{Spectrum of bosonic excitations in $SU(N)$ gauge
  theory for $N_f$ close to $N^c_f$. For $N_f < N^c_f$, chiral symmetry
  is broken.  For $N_f \stackrel{>}{\sim} N^c_f$, the theory is assumed
  to be in a conformally-invariant ``non-abelian coulomb'' (NAC) phase [2]
  which has no isolated single-particle states.}
\label{Fig2}
\end{minipage}
\end{figure}

This picture should be contrasted with the analogous change of the
spectrum of particles near the chiral phase transition in non-abelian
gauge theory as $N_f \to (N_f^c)^-$. Because of the infrared
fixed-point, the high-energy scale $\Lambda$ (see eqn. \ref{critical})
is no longer relevant.  In this case the {\it only} relevant dynamical
scale is the magnitude of dynamical mass $\Sigma(0)$.  All other scales,
the $F$-constant, the confinement scale, {\it etc.} are of the same
order of magnitude \cite{atw}.  Therefore, if $\Sigma(0) \to 0$
continuously as $N_f \to (N_f^c)^-$, we expect the spectrum of bosonic
excitations to be as shown in Figure 2.  Note that as $N_f \to
(N_f^c)^-$, the entire spectrum collapses to zero mass. So long as the
Banks-Zaks fixed point persists\footnote{More properly, since we are in
  the broken phase where no true fixed-point exists, so long as the
  coupling remains close to $\alpha^*$ over a large range of momenta.}
in the non-abelian gauge theory, all high-energy scales are irrelevant.
Therefore, unlike the NJL model, if $\Sigma(0) \to 0$ the mass of all
excitations must also tend to zero.

What are the implications of this behavior? In the case of the NJL model
near the critical coupling, it is appropriate to ``integrate-out'' all
higher-mass states and the critical behavior of the theory is determined
entirely by the infrared behavior of the corresponding scalar field
theory\footnote{In some cases the resulting scalar theory cannot have a
  second order transition, but will have a fluctuation-induced
  first-order transition instead \cite{cw}. If the transition is driven
  to be first-order, all relevant dimensional quantities will be
  \cite{cgs, bardeen} of order $\Lambda$.}. We therefore expect similar
behavior in {\it any} $SU(N_f) \times SU(N_f)$ chirally-invariant
four-dimensional field theory with {\it short-range interactions} near the
phase boundary: {\it i.e.} we expect the behavior to be {\it universal}.

In contrast, in the case of non-abelian gauge theory with $N_f$ just
below $N_f^c$, we {\it cannot} reduce the theory to an effective
low-energy scalar theory. In principle {\it all} ``higher'' resonances
contribute to any correlation function and the behavior is {\it
  non-universal}.  In this case it is plausible that the order parameter
changes continuously as $N_f \to (N_f^c)^-$, even though the theory with
$N_f \stackrel{>}{\sim} N^c_f$ has no light scalar states.

Finally, I note that arguments similar to those presented here may cast
light on the discrepancies between the gap-equation \cite{QED3, QED3PT}
and the effective scalar field-theory \cite{pisarski} analyses of the
chiral phase transition in QED3.

\bigskip

In this note I have proposed a simple physical picture for the spectrum
of states in $SU(N)$ gauge theory as the number of flavors $N_f$
approaches the critical number $ N^c_f$ from below ({\it i.e.} from the
broken side). This picture predicts very different and {\it
  non-universal} behavior than is the case in a conventional second
order phase transition. I have further argued that this feature results
from the (presumed) existence of a Banks-Zaks fixed point of the gauge
coupling in the neighborhood of the chiral transition and therefore
depends on the long-distance nature of the non-abelian gauge force.

\bigskip


\centerline{\bf Acknowledgments}

I thank Koichi Yamawaki and the organizers of the 1996 International
Workshop on Perspectives of Strong Coupling Gauge Theories (SCGT 96)
held in Nagoya, Japan from 13-16 November, 1996 for holding a
stimulating conference where this work was begun.  I also thank Nick
Evans, Stephen Selipsky, Rohana Wijewardhana, Elizabeth Simmons and John
Terning for discussions and for comments on the manuscript. {\em This
  work was supported in part by the Department of Energy under grant
  DE-FG02-91ER40676.}


\vfill\eject

\end{document}